# Removing krypton from xenon by cryogenic distillation to the ppq level

**XENON Collaboration**[1,*]

E. Aprile[2], J. Aalbers[3], F. Agostini[4,5], M. Alfonsi[6], F. D. Amaro[7], M. Anthony[2], F. Arneodo[8], P. Barrow[9], L. Baudis[9], B. Bauermeister[6,10], M. L. Benabderrahmane[8], T. Berger[11], P. A. Breur[3], A. Brown[3], E. Brown[11], S. Bruenner[12], G. Bruno[4], R. Budnik[13], L. Bütikofer[14,a], J. Calvén[10], J. M. R. Cardoso[7], M. Cervantes[15], D. Cichon[12], D. Coderre[14,a], A. P. Colijn[3], J. Conrad[10,b], J. P. Cussonneau[16], M. P. Decowski[3], P. de Perio[2], P. Di Gangi[5], A. Di Giovanni[8], S. Diglio[16], E. Duchovni[13], G. Eurin[12], J. Fei[17], A. D. Ferella[10], A. Fieguth[18], D. Franco[9], W. Fulgione[4,19], A. Gallo Rosso[4], M. Galloway[9], F. Gao[2], M. Garbini[5], C. Geis[6], L. W. Goetzke[2], L. Grandi[20], Z. Greene[2], C. Grignon[6], C. Hasterok[12], E. Hogenbirk[3], C. Huhmann[18], R. Itay[13], B. Kaminsky[14,a], G. Kessler[9], A. Kish[9], H. Landsman[13], R. F. Lang[15], D. Lellouch[13], L. Levinson[13], M. Le Calloch[16], Q. Lin[1], S. Lindemann[12], M. Lindner[12], J. A. M. Lopes[7,c], A. Manfredini[13], I. Maris[8], T. Marrodán Undagoitia[12], J. Masbou[16], F. V. Massoli[5], D. Masson[15], D. Mayani[9], Y. Meng[21], M. Messina[2], K. Micheneau[16], B. Miguez[19], A. Molinario[4], M. Murra[18,d], J. Naganoma[22], K. Ni[17], U. Oberlack[6], S. E. A. Orrigo[7,e], P. Pakarha[9], B. Pelssers[10], R. Persiani[16], F. Piastra[9], J. Pienaar[15], M.-C. Piro[11], V. Pizzella[12], G. Plante[2], N. Priel[13], L. Rauch[12], S. Reichard[15], C. Reuter[15], A. Rizzo[2], S. Rosendahl[18,f,g], N. Rupp[12], R. Saldanha[20], J. M. F. dos Santos[7], G. Sartorelli[5], M. Scheibelhut[6], S. Schindler[6], J. Schreiner[12], M. Schumann[14], L. Scotto Lavina[23], M. Selvi[5], P. Shagin[22], E. Shockley[20], M. Silva[7], H. Simgen[12], M. v. Sivers[14,a], A. Stein[21], D. Thers[16], A. Tiseni[3], G. Trinchero[19], C. Tunnell[3,20], N. Upole[20], H. Wang[21], Y. Wei[9], C. Weinheimer[18], J. Wulf[9], J. Ye[17], Y. Zhang[2], and I. Cristescu[24]

[1] Laboratori Nazionali del Gran Sasso, Assergi, Italy
[2] Physics Department, Columbia University, New York, NY, USA
[3] Nikhef and the University of Amsterdam, Science Park, Amsterdam, Netherlands
[4] INFN-Laboratori Nazionali del Gran Sasso and Gran Sasso Science Institute, L'Aquila, Italy
[5] Department of Physics and Astrophysics, University of Bologna and INFN-Bologna, Bologna, Italy
[6] Institut für Physik & Exzellenzcluster PRISMA, Johannes Gutenberg-Universität Mainz, Mainz, Germany
[7] Department of Physics, University of Coimbra, Coimbra, Portugal
[8] New York University Abu Dhabi, Abu Dhabi, United Arab Emirates
[9] Physik-Institut, University of Zurich, Zurich, Switzerland
[10] Oskar Klein Centre, Department of Physics, Stockholm University, AlbaNova, Stockholm, Sweden
[11] Department of Physics, Applied Physics and Astronomy, Rensselaer Polytechnic Institute, Troy, NY, USA
[12] Max-Planck-Institut für Kernphysik, Heidelberg, Germany
[13] Department of Particle Physics and Astrophysics, Weizmann Institute of Science, Rehovot, Israel
[14] Physikalisches Institut, Universität Freiburg, 79104 Freiburg, Germany
[15] Department of Physics and Astronomy, Purdue University, West Lafayette, IN, USA
[16] SUBATECH, Ecole des Mines de Nantes, CNRS/In2p3, Université de Nantes, Nantes, France
[17] Department of Physics, University of California, San Diego, CA, USA
[18] Institut für Kernphysik, Westfälische Wilhelms-Universität Münster, Münster, Germany
[19] INFN-Torino and Osservatorio Astrofisico di Torino, Torino, Italy
[20] Department of Physics and Kavli Institute of Cosmological Physics, University of Chicago, Chicago, IL, USA
[21] Physics and Astronomy Department, University of California, Los Angeles, CA, USA
[22] Department of Physics and Astronomy, Rice University, Houston, TX, USA
[23] LPNHE, Université Pierre et Marie Curie, Université Paris Diderot, CNRS/IN2P3, Paris 75252, France
[24] Tritium Laboratory Karlsruhe, Karlsruhe Institute of Technology, Eggenstein-Leopoldshafen, Germany

I. Cristescu: Not member of the XENON Collaboration
* e-mail: xenon@lngs.infn.it

[a] Also with Albert Einstein Center for Fundamental Physics, University of Bern, Switzerland

Springer

**Abstract** The XENON1T experiment aims for the direct detection of dark matter in a detector filled with 3.3 tons of liquid xenon. In order to achieve the desired sensitivity, the background induced by radioactive decays inside the detector has to be sufficiently low. One major contributor is the $\beta$-emitter $^{85}$Kr which is present in the xenon. For XENON1T a concentration of natural krypton in xenon $^{nat}$Kr/Xe $<$ 200 ppq (parts per quadrillion, 1 ppq $= 10^{-15}$ mol/mol) is required. In this work, the design, construction and test of a novel cryogenic distillation column using the common McCabe–Thiele approach is described. The system demonstrated a krypton reduction factor of $6.4 \cdot 10^5$ with thermodynamic stability at process speeds above 3 kg/h. The resulting concentration of $^{nat}$Kr/Xe $<$ 26 ppq is the lowest ever achieved, almost one order of magnitude below the requirements for XENON1T and even sufficient for future dark matter experiments using liquid xenon, such as XENONnT and DARWIN.

## 1 Introduction

With a view to discover a possible candidate of the elusive dark matter – namely, the weakly interacting massive particle (WIMP) [1–3] – the XENON dark matter project for direct WIMP detection has been launched. Using a xenon dual-phase time projection chamber (TPC), it searches for an energy deposition in the detector by the elastic scattering of WIMPs off xenon nuclei. The XENON1T detector with a total amount of liquid xenon of 3.3 tons has started data taking and is designed to reach a sensitivity more than an order of magnitude beyond the current limits with a two ton-year exposure, probing new dark matter parameter space [4].

The intrinsic contamination of the xenon itself is one of the most critical contributors to the total background for large detector masses, and thus for the sensitivity of the experiment. In the calculation of the projected sensitivity, the electronic recoil background was assumed to be equal to $1.80 \cdot 10^{-4}$ (kg · day · keV)$^{-1}$, mainly due to radon intrinsic to the liquid xenon [4]. Another key contributor is $^{85}$Kr, a $\beta$-emitter with a half-life of 10.76 y and an endpoint energy of 687 keV. Although background reduction techniques are applied at the analysis level to discriminate these beta decays from WIMP-induced nuclear recoils, active removal of krypton is needed before starting the dark matter search to reduce the remaining background to a tolerable level. For the XENON1T experiment, a concentration of natural krypton in xenon $^{nat}$Kr/Xe $<$ 200 ppq,[1] assuming a ratio of $^{85}$Kr/$^{nat}$Kr $= 2 \cdot 10^{-11}$, is necessary to keep the background from krypton subdominant with respect to the one from radon, and so to meet the background requirements [4]. In future experiments, using multiple tonnes of liquid xenon as fiducial target, the purity constraints are even more stringent, e.g., for the XENONnT [4], LZ [5] and DARWIN [6] experiments, where concentrations well below 100 ppq have to be achieved. Commercially available xenon gas can be delivered with krypton concentrations in the range from 1 ppm[2] to 10 ppb,[3] which is generally not clean enough for dark matter experiments using liquid xenon detectors.

Separation of krypton from xenon can be achieved thanks to the different physical properties of xenon and krypton, such as the difference in vapour pressures or their different mobilities in porous media. The former property is exploited in cryogenic distillation while the latter is used in chromatographic systems. The LUX experiment uses gas chromatography to achieve a krypton concentration in xenon down to 3.5 ppt[4] after a single-pass through the removal system and below 200 ppq after two passes at a purification speed of 0.4 kg/h [7]. Cryogenic distillation has first been utilized by XMASS, reaching a concentration of 3 ppt [8]. The same type of distillation column was used in XENON100 to produce a purified output concentration of 19 ppt at a purification speed of 0.6 kg/h [9]. After several cycles of distillation, even lower concentrations of about 1 ppt were obtained in XENON100 [10] and PandaX [11].

In this paper, we present a new distillation column, designed and built with an increased purification speed of 3 kg/h [8.3 standard liters per minute (slpm)]. It features a minimal loss of krypton-enriched xenon of 1% and a higher separation performance. Starting with a concentration in the ppb range, a reduction factor of $F_{\text{red}} \approx 10^4 - 10^5$ is needed to achieve the purity required for XENON1T. The reduction factor is defined as the ratio of the krypton concentration in the in-gas $c_F$ divided by the concentration in the purified liquid-out sample $c_B$

$$F_{\text{red}} = \frac{^{nat}\text{Kr/Xe}_{(\text{in-gas})}}{^{nat}\text{Kr/Xe}_{(\text{liquid-out})}} \equiv \frac{c_F}{c_B}. \quad (1)$$

Section 2 introduces the conceptional design of the new system based on the commonly used McCabe–Thiele method [12]. The detection of krypton in xenon is an important aspect when characterizing distillation systems, which is treated in Sect. 3. The development of the distillation column was divided into three phases. It started with a single stage distillation setup (Phase-0), already presented in [13], before starting the construction of a multi-stage test setup (Phase-1) to investigate different aspects of the separation, summarized in Sect. 4. The final, and improved, setup for the XENON1T experiment (Phase-2) was commissioned at the Laboratori Nazionali del Gran Sasso (LNGS) in Italy and is presented in Sect. 5.

---

$^b$ Wallenberg Academy Fellow
$^c$ Also with Coimbra Engineering Institute, Coimbra, Portugal
$^d$ e-mail: michaelmurra@wwu.de
$^e$ Now at IFIC, CSIC-Universidad de Valencia, Valencia, Spain
$^f$ Now at Physikalisch-Technische Bundesanstalt (PTB), Braunschweig, Germany
$^g$ e-mail: stephan.rosendahl@ptb.de

---

[1] Parts per quadrillion, 1 ppq $= 1 \cdot 10^{-15}$ mol/mol.
[2] Parts per million, 1 ppm $= 1 \cdot 10^{-6}$ mol/mol.
[3] Parts per billion, 1 ppb $= 1 \cdot 10^{-9}$ mol/mol.
[4] Parts per trillion, 1 ppt $= 1 \cdot 10^{-12}$ mol/mol.

## 2 Design of the cryogenic distillation column

The distillation process as described in [12,14] is based on the difference of vapour pressures of the two components in an ideal binary mixture, in our case krypton and xenon. Assuming a static liquid xenon reservoir in equilibrium with the gaseous phase above, the more volatile component (krypton) is enriched in the gaseous phase. The enhancement in concentration can be described by the relative volatility $\alpha$, which is deduced by using Raoult's law to be the ratio between the vapour pressures of the two components:

$$\alpha = \frac{P_{Kr}}{P_{Xe}} = 10.8 \quad (\text{at T} = -98°C) \quad [15]. \tag{2}$$

The volatility can also be interpreted as a measure of the probability for krypton to migrate into the gaseous phase at liquid xenon temperatures.

In Phase-0, a single stage distillation system was realized, containing liquid xenon with a gaseous phase above. Such a system allowed the investigation of the dynamics of a distillation process in a volume with a single liquid-gas interface and steady state mass flow at very low concentrations of krypton in xenon in the ppq range [13]. This allowed the first measurement of the relative volatility at such low concentrations of krypton in xenon, finding that $\alpha \approx 10$, consistent with Raoult's law even in this regime. The measurements also confirmed the general prediction that a single distillation stage is not enough to achieve a separation of $10^4$–$10^5$ [14]. Therefore, the distillation column presented here employs the techniques commonly used in distillation plants, where a series of distillation stages is used to successively reduce the krypton concentration to the desired level of purity.

In the simplest configuration, the distillation column is treated as a system with three sections as shown in Fig. 1. The feeding section where the xenon is injected to the distillation tower, the rectifying section above where the more volatile component is enriched, and the stripping section below the feeding section where the more volatile component is depleted. Setting the reboiler stage in the stripping section as the starting point, the binary mixture is evaporated and a gaseous stream flows upwards at a flow rate of $V'$. Assuming that the vapour will condense on the next stage (Plate 1),

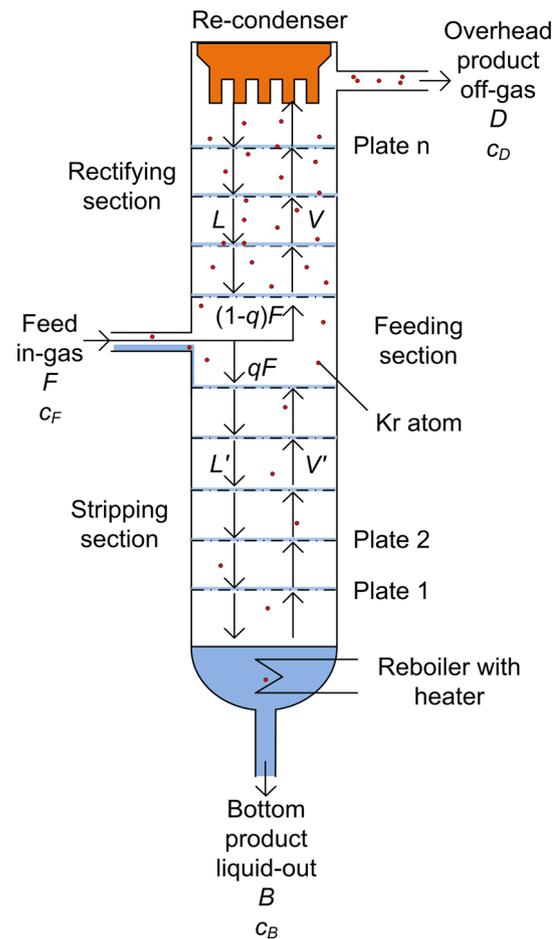

**Fig. 1** Schematics of a distillation column with partial reflux. In this diagram, the typical construction of a distillation column using several distillation plates and partial reflux is shown. It consists of three parts, the stripping section, the feeding section and the rectifying section. $L$ and $L'$ are the down-going liquid streams while $V$ and $V'$ are the up-going gaseous streams in the different parts. Both are connected to the feeding stream $F$. The krypton enriched off-gas stream $D$ is extracted from the top while the purified xenon is extracted from the bottom with a flow rate $B$. The krypton in xenon concentrations of the feed and the two extraction flows are given by $c_F$, $c_D$ and $c_B$, while $q$ defines the state of matter of the injected xenon

the generated liquid phase on this stage will have the same composition as the vapour from the reboiler stage, and will therefore have a higher concentration of krypton compared to the starting mixture. By repeating this procedure for several stages, the fraction of krypton in xenon will successively increase. On top of the tower, the vapour is partially liquefied with the help of the re-condenser and fed back to the column with a flow rate $L$. A small fraction of the gas is taken out as krypton enriched off-gas at a flow rate $D$, while the purified xenon is extracted at the bottom with a flow rate $B$. Such an operation with a partial reflux is called rectification.

The xenon, injected at the feeding section at a flow rate $F$, is added to either the up-streaming vapour $V$ in the rectifying section or to the down-streaming liquid flow $L'$. This depends on the state of the injected matter (gaseous or liquid), which is described by the caloric factor $q = (L' - L)/F$. The design for the XENON1T column foresees the usage of a saturated liquid feed at the boiling point ($q = 1$). Depending on the requested krypton in xenon concentration at the bottom (purified xenon, liquid-out) and at the top (krypton enriched xenon, off-gas) the number of distillation stages can be calculated using the McCabe–Thiele method [12]. Although the validity of this method has never been demonstrated at such low concentrations, we followed it to design our distillation column. We will later compare our findings with the predictions. In this method the concentration of krypton in the gaseous and liquid phase can be calculated to derive an equilibrium diagram, which is shown in Fig. 2.

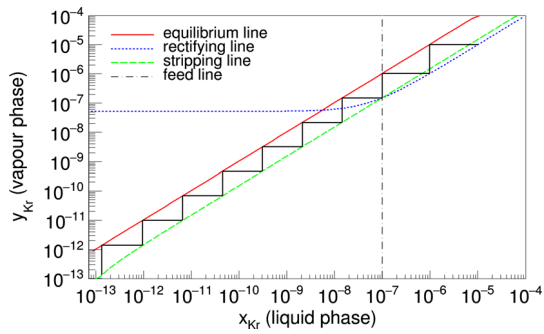

**Fig. 2** McCabe–Thiele diagram for the XENON1T requirements. The concentration of the krypton in the gas phase $y_{Kr}$ in relation to the concentration in the liquid phase $x_{Kr}$. The operation lines for the different sections, namely rectifying (*blue dotted line*), stripping (*green dashed line*) and feed line (*grey dash-dotted line*) of the column are drawn and allow for the determination of the number of theoretical stages needed to meet the XENON1T requirements (*black solid line*). The definitions of the equilibrium (*red solid line*) and operation lines are taken from [12,14]

Based on the vapour pressures and the relative volatility $\alpha$, the relation between the krypton concentration in the gaseous phase $y_{Kr}$ and the liquid phase $x_{Kr} = 1 - x_{Xe}$ for very low concentrations ($x_{Kr} \cdot (\alpha - 1) \ll 1$) in equilibrium is defined as

$$y_{Kr} = \frac{P_{Kr} \cdot x_{Kr}}{P_{Kr} \cdot x_{Kr} + P_{Xe} \cdot x_{Xe}} \tag{3}$$

$$= \frac{\alpha \cdot x_{Kr}}{1 + (\alpha - 1) \cdot x_{Kr}} \tag{4}$$

$$\approx \alpha \cdot x_{Kr}. \tag{5}$$

This relation is called the equilibrium line and is shown in red in Fig. 2.

By heating from the bottom and cooling from the top, the equilibrium inside the column is disturbed by the upward gaseous and downward liquid flows introduced in the column. The actual concentration of krypton in the gaseous phase related to the concentration in the liquid phase can be described, due to the mass flow, by operating lines, defined below. For each section, an operation line for the diagram is determined, shown in Fig. 2 marked as gray, blue and green lines respectively. A forcing concentration gradient is ensured between the equilibrium line and operating lines, driving the system towards the equilibrium via mass transfer between the ascending gaseous stream and the descending liquid stream.

The determination of the operation lines is briefly reviewed here for the example of the rectifying section. For a complete derivation of the formulas on the McCabe Thiele method we would like to refer to the literature, e.g. [12,14]. The total mass balance depends on the design specifications as introduced in Sect. 1. The incoming flow $F = 3$ kg/h (in-gas) is equal to the outgoing flows at the reboiler $B = 2.97$ kg/h (purified xenon, liquid-out) and at the condenser $D = 0.03$ kg/h (krypton enriched xenon, off-gas, fixed by design to 1% of in-gas) during stable operation, giving the total mass-balance,

$$F = D + B. \tag{6}$$

This can also be written in terms of krypton atoms, using the different concentrations of krypton in the feed $c_F$, in the off-gas $c_D$ and in the liquid-out $c_B$,

$$F \cdot c_F = D \cdot c_D + B \cdot c_B. \tag{7}$$

For an input concentration of $c_F = 100$ ppb and the pre-defined purified output concentration for XENON1T, $c_B = 200$ ppq, the concentration in the off-gas can be calculated as $c_D = 10$ ppm for the given flows.

In the rectifying section, the krypton concentration on a distillation stage $y_{Kr}$ in the vapour stream $V$ can be calculated using the krypton particle balance to get the operating line (rectifying line)

$$V \cdot y_{Kr} = L \cdot x_{Kr} + D \cdot c_D \tag{8}$$

$$\rightarrow y_{Kr} = \frac{R}{R+1} \cdot x_{Kr} + \frac{c_D}{R+1}, \tag{9}$$

where $L$ is the liquid xenon flux in the rectifying section. $R$ is called the reflux ratio and describes the amount of xenon that is fed back as liquid ($L$) from the top, related to the flow of extracted off-gas $D$

$$R = \frac{L}{D}. \tag{10}$$

For good separation performance, a high reflux ratio $R$ is mandatory [12,14]. Since $R = 191$ worked well for the XENON100 column [8] and was not the limiting factor in terms of separation efficiency, the same value for the XENON1T system was used in the initial design. The liquid xenon flux in the rectifying section is calculated as $L = R \cdot D = 5.73$ kg/h, which directly defines the amount of cooling power needed at the top condenser to be $P_{\text{Top}} = 147.4$ W in order to liquefy the xenon.

The McCabe–Thiele diagram, shown in Fig. 2, was drawn for a saturated liquid feed ($q = 1$), indicated by the intersection line (in this case vertical) of the two operation lines and a starting concentration of $c_F = 100$ ppb. The number of black steps in the diagram indicates the number of theoretical stages and can be determined by taking the liquefaction processes and the related vapour compositions into account. A liquid mixture with the krypton content $x_{\text{Kr}}$ has a vapour phase with an enhanced krypton concentration $y_{\text{Kr}}$, which is streaming up to the next stage (vertical line) where it gets liquefied again (horizontal line).

In total, nine theoretical distillation stages are needed to reach a reduction factor of $F_{\text{red}} \sim 10^4$–$10^5$. This allows xenon with the highest commercially available purity ($\sim$ppm) to be purified in a single pass through the column. For a saturated vapour feed ($q = 0$), the diagram changes slightly and 10 stages are needed in order to achieve the same separation.

In practice, multiple single distillation stages are not practical. However, by utilizing a structured package material inside the column, featuring a large surface area for good liquid-gas exchange through the full height of the column, the effective exchange is equivalent to a certain number of theoretical single distillation stages. The height-equivalent of one theoretical plate[5] (HETP) denotes the amount of required package material to realize one distillation stage. Together with the number of predicted plates $N_{\text{TP}}$ derived from the McCabe–Thiele diagram, the total height $h$ of the package tube can be calculated:

$$h = \text{HETP} \cdot N_{\text{TP}}. \quad (11)$$

The HETP value is determined empirically and varies for different compositions and system parameters. Typical values for the package type used here are (2–8) cm [16], but it has not been measured for a krypton-xenon mixture so far. Furthermore, the partial pressure of krypton in the injected krypton-xenon mixture is already very low, due to its low feed concentration $c_F \sim 100$ ppb. The investigations on the single stage distillation setup confirmed the separation process on the ppq scale for krypton and xenon to be in agreement with the expectations from the relative volatility [13]. However, for this concentration regime the McCabe–Thiele approach has not been validated for larger distillation plants. Due to these uncertainties it was decided to first realize a smaller distillation column, using 1 m of package material (Phase-1 column) and increase the amount of material afterwards, if needed (Phase-2 column). The components of the Phase-1 column were designed to allow an easy upgrade from Phase-1 to Phase-2.

---

[5] One theoretical plate equals one distillation stage.

## 3 Diagnostics of krypton in xenon

The measurement of trace amounts of krypton in xenon is important for the characterization of the cryogenic distillation column. The XENON collaboration developed several techniques for trace gas measurements to detect natural krypton in xenon to the sub-ppt level [10,17,18]. In addition, a $^{83m}$Kr tracer method was developed where $^{83m}$Kr with a half-life of 1.83 h is injected to the feed gas. The measurement of the decay of $^{83m}$Kr, which chemically behaves as natural krypton, at the inlet and the two outlets of the column allows investigating the stability of the distillation process as well as its dynamics over time. The method has been successfully used for the single stage system as well as for the Phase-1 and Phase-2 columns. It is presented in more detail in [13,19].

### 3.1 RGA with coldtrap enhanced sensitivity

For an online measurement of the absolute concentration of natural krypton in xenon, a system based on [20,21] was developed at Westfälische Wilhelms-Universität (WWU) Münster. By using a residual gas analyzer (RGA) behind a LN$_2$-cooled coldtrap, an enhanced sensitivity for concentrations down to the sub-ppb level was achieved. The xenon gas sample ($\approx$90 ml) was introduced into a spiral gas routing pipe, which was cooled to 77 K by submerging it in liquid nitrogen. This provides an artificial enhancement of the krypton concentration as most of the xenon is frozen in the coldtrap whereas the krypton remains mostly gaseous due to its higher vapour pressure and low partial pressure. The sample was injected into a measurement chamber equipped with a commercial quadrupol mass-filter (Inficon, Transpector 2 H200M) and an oil-free turbo-molecular pumping system. A custom-made butterfly valve was mounted before the pump, which allowed the control of the gas load inside the measurement chamber by adjusting the effective pumping speed. This increased the dynamic range and provided higher sensitivity. A similar version described in [18] was constructed with a sensitivity limit of (206 ± 36) ppt, obtained from calibration with different known krypton concentrations.

Since the expected concentration of the distilled xenon is at the ppq level, only a lower limit for the separation efficiency can be given. However, this setup provides valuable online feedback on the performance of the distillation column since it is connected to the inlet and all outlets of the distillation system.

### 3.2 GC-RGMS - Gas chromatography in combination with residual gas mass spectrometry

The most sensitive device used to characterize the Kr-removal capability of the distillation columns is the rare gas mass spectrometer method (RGMS), developed and located

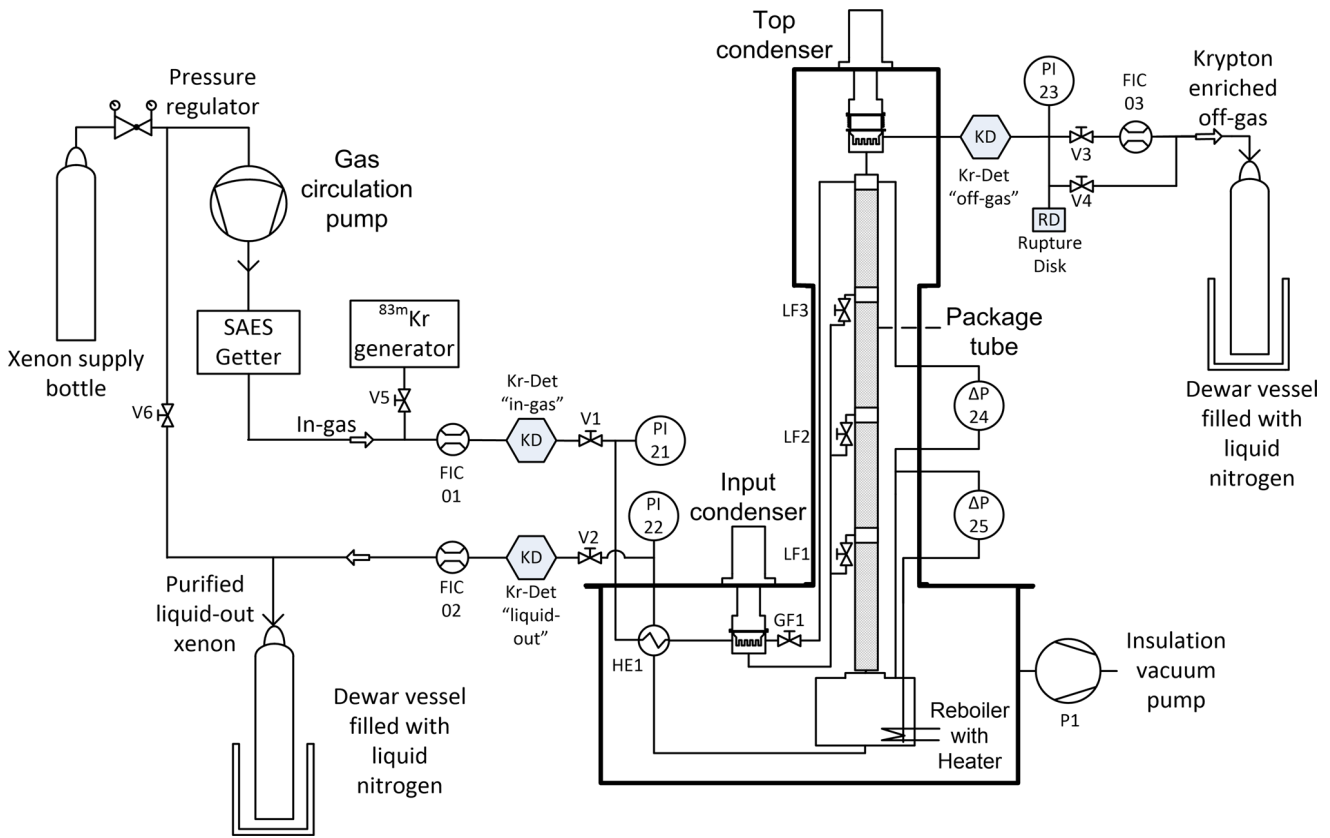

**Fig. 3** P & I Diagram of the Phase-1 distillation setup. The xenon is injected from storage bottles to the distillation plant, while the purified xenon as well as the krypton enriched off-gas are collected into separate gas bottles by cryo-pumping. The scheme is simplified and shows only the most important sensors for pressure and flow rates which are required for the regulation

at the Max-Planck-Institut für Kernphysik in Heidelberg. It detects trace amounts of natural krypton in xenon down to the ppq level. The projected detection limit was determined to be $^{nat}Kr/Xe = 8\,ppq$ [10]. Similar to the coldtrap RGA technique, a separation of the bulk of xenon from the krypton component has to be done before the sample is injected into the analyzer. This separation is realized by gas-chromatography.

## 4 The Phase-1 distillation column

### 4.1 Design and assembly

The Phase-1 distillation column has a total height of 3.5 m and uses 1.1 m of package material in the distillation tower. Figure 3 shows the piping and instrumentation diagram (P & ID) of the system as well as the infrastructure which allows for different performance studies. The column consists of four main components, the input condenser, the package tube, the top condenser and the reboiler. At the input condenser, equipped with a cryo-cooler (Leybold CP-50), the injected xenon is liquefied with a maximum cooling power of 100 W at $-98\,°C$, before it is passed to the package tube where the distillation process takes place. The tube is filled with structured package material made from stainless steel (Sulzer EX, diameter 45 mm) as shown in Fig. 4 and is equipped with different feeding ports (labeled LF1-LF3 for liquid feed injected from the bottom and GF1 for gaseous feed injected from the top of the input condenser) in order to allow for different feeding conditions. The top condenser is equipped with a cryo-cooler (Leybold CP-140T) providing 200 W of cooling power at $-98\,°C$. The reboiler at the bottom stores and evaporates the liquid xenon using heater cartridges with a maximum of 300 W heating power available. The entire setup is housed inside a vacuum insulation vessel, pumped to $10^{-5}$ mbar in order to minimize thermal losses.

Both condensers are equipped with calibrated silicon diodes (Lakeshore DT-670D-CU) providing accurate temperature monitoring, while additional Pt-1000 sensors at different locations complete the overall temperature supervision (not shown in Fig. 3). The most important system parameter is the pressure inside the package tube, measured with a capacitive manometer (MKS Baratron 121A, labeled as PI23) while the inlet and the outlet pressures PI21 and PI22 are measured with a capacitive manometer with lower precision (Swagelok PTU series). The differential pressure along the package tube is also measured (MKS Baratron 221B; $\Delta$P24). The same type of differential pressure sensor ($\Delta$P25) is used to obtain the height of the liquid xenon level inside the reboiler by measuring the hydrostatic pressure of the liquid in relation to the column pressure.

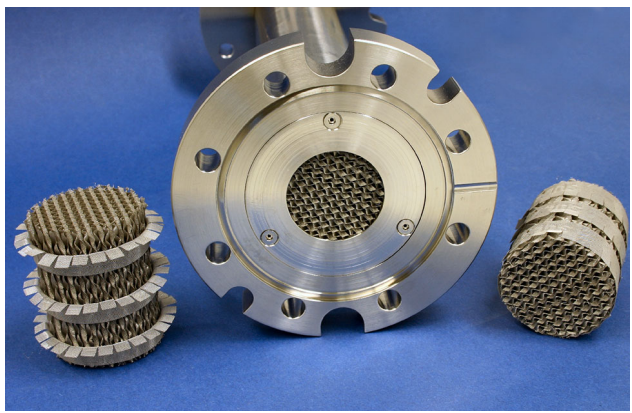

**Fig. 4** Package tube. After a cleaning procedure, the structured package material is filled into the package tube inside a clean-room

The mass-balance inside the system is controlled by three mass-flow controllers (MKS 1479B) labeled as FIC01 for the in-gas, FIC02 for the purified liquid-out and FIC03 for the krypton enriched off-gas. The liquid-out and the in-gas flows are thermodynamically coupled by a plate-heat exchanger (GEA GBS 100M-200R; HE1). A rupture disc (RD) was added for the release of accidental overpressures greater than 4 bar. Three $^{83m}$Kr decay detectors (KD) are implemented at the in-gas, as well as at the liquid-out and the off-gas lines [19].

In order to guarantee a high xenon purity, all parts were either made from electro-polished stainless steel (alloy 316L and 316LN) or from oxygen-free, high thermal conductivity (OFHC) copper. All connections were either made by metal sealed VCR connections or CF flanges, while the fittings for the VCR connection were orbitally welded to the pipes to achieve optimal leak-tightness and clean weld seams. All components were very carefully cleaned and the assembled system was baked, evacuated and xenon gas was circulated through a hot getter (SAES PS4-MT15-R-2) in a closed loop via V6 to further reduce remaining impurities.

### 4.2 Commissioning

Before starting the distillation process, the column is cooled down and filled with xenon in order to achieve a certain liquid level inside the reboiler vessel, (2–3) cm corresponding to (3–4) kg, and the input condenser (1 kg) to realize the liquid feed condition. The regulation circuit stabilizes the pressure inside the system to about 2.0 bar absolute by adjusting the heater power in the reboiler and by keeping the applied cooling power at the top condenser constant, thereby establishing a stable xenon flux along the package tube. The distillation run is performed by flushing xenon from the storage bottle, through the getter purification stage, and into the distillation column. For the storage of purified liquid-out gas, and also for krypton enriched off-gas, two aluminum gas bottles are cooled in a liquid nitrogen bath.

### 4.3 Thermodynamic stability

For an optimal separation of krypton and xenon, the distillation process needs to be thermodynamically stable such that the mass exchange between liquid and gas streams are guaranteed. Thus, it is working ideally after the pressure in the system has been stabilized and mass balance is achieved. This means that the amount of xenon gas that is fed into the column equals to the extracted xenon gas. Thus, the flows are adjusted until the balance $F = B + D$ is reached. Figure 5 presents relevant parameters of an 11 h distillation run, using a liquid feed condition ($q = 1$), injected at port LF2 (see Fig. 3). The first diagram shows the absolute pressure inside the distillation column (PI23), which was set to 1.9 bar. After five hours of regulation time, the measured pressure at the sensor was PI23 = 1.90 bar, matching the set-point with a standard deviation of $\sigma_{PI23} = 0.006$ bar. The differential pressure along the package tube ($\Delta$P24) is an important parameter in order to avoid flooding of the package with liquid xenon which can reduce the performance. The differential pressure was measured to be $\Delta$P24 = (1.0 ± 0.1) mbar leading to a pressure drop of about 0.91 mbar/m of package material, slightly below the recommended value of 1 mbar/m [16]. Consequently, there is no flooding of the package during this run.

The differential pressure in the reboiler indicates that the mass-balance and the feeding is working properly. When the distillation is started, the liquid level changes due to feeding, liquid extraction at the bottom, and turbulences due to the regulation. After some time it stabilizes at $\Delta$P25 = (6.0 ± 0.1) mbar (equivalent to a height of 22 mm). The relative change in the liquid level during the last 5 h of the distillation was only −0.2%, indicating very good mass-balance stability. Finally, the third panel shows the different flow rates at the in-gas and the purified liquid-out lines. The run was performed at an average flow rate of FIC01 = 8.5 slpm, with a fraction of 1.1% going into the off-gas. The reflux ratio was measured to be $R_{Meas} = 183$, slightly smaller than foreseen from Sect. 2. This can be explained by heat-input to the system from the outside, by thermal radiation or heat bridges, leading to a reduced effective cooling power at the top condenser and thus a smaller reflux. Nevertheless, the effect on the performance is expected to be small according to McCabe–Thiele diagram calculations.

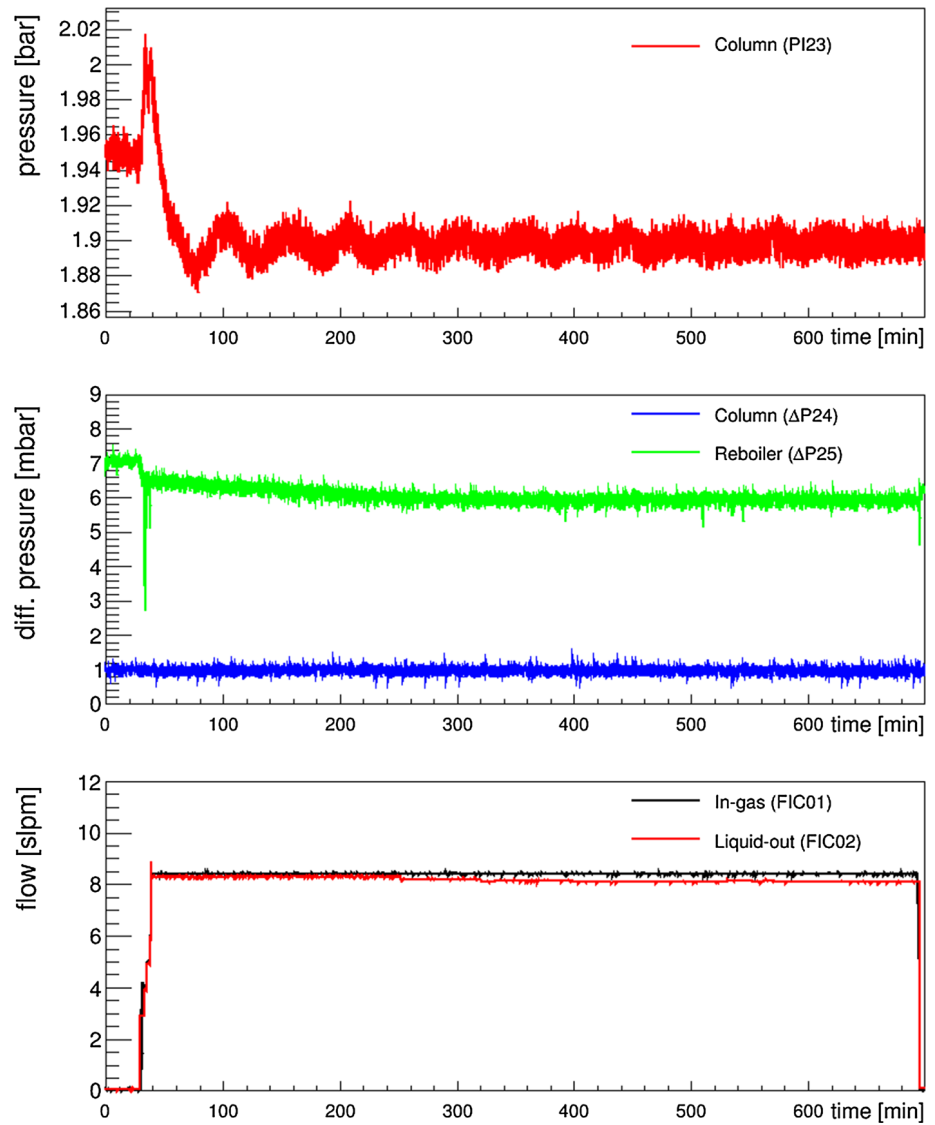

**Fig. 5** Thermodynamic stability of the distillation process. Shown are the key parameters during a distillation run of about 11 h. The width and small spikes of the measurements are created by sensor inaccuracy and do not reflect the actual performance. *Top* Pressure PI23 inside the package material. *Middle* Pressure drop along the package tube ($\Delta$P24) and hydrostatic pressure of the liquid xenon reservoir in the reboiler ($\Delta$P25). *Bottom* Flow rates in the in-gas line (FIC01) and in the liquid-out line (FIC02)

Additional performance studies were carried out at increased process speeds of up to 16 slpm which turned out to be thermodynamically stable. These rates could only be achieved with the heat exchanger HE1, working at an efficiency of >90%, in agreement with measurements in [22]. Without the heat exchanger, the cooling power provided by the input condenser would not be sufficient to guarantee a saturated liquid feed. The final limit on the process speed for the Phase-1 system is given by the flow controllers which are capable of a maximum speed of 20 slpm (=7.2 kg/h).

### 4.4 Separation efficiency

Since the purified output was well below the sensitivity limit of the coldtrap RGA system, the RGMS technique was used to measure the reduction factor of the distillation column. A xenon gas sample was extracted into a transport container and shipped to MPIK Heidelberg. In order to avoid any contamination of the sample, the container consisted of four metal-sealed 1/2″ VCR valves, welded together in series to enclose three volumes of a total of about 30 cm$^3$ which were filled with sample gas. The main advantage of three volumes is the separation of the central volume by the two adjacent ones from ambient air which contains a 1.1 ppm fraction of natural krypton. In this way the central volume is well protected and potential tiny leaks in the valves would show up in the two adjacent volumes first. Finally, two containers were used to draw samples from the in-gas and the purified liquid-out lines. They were mounted to the Phase-1 column, baked and evacuated for two days to establish a clean environment inside the storage container before the samples were taken.

**Table 1** Results from the concentration measurement of the purified liquid-out sample. Six measurements using the RGMS system were performed on the three volumes containing the liquid-out sample. The sample was drawn during the commissioning of the Phase-1 column

| Volume | Xe [cm$^3$] | $^{nat}$Kr/Xe [ppq] |
| --- | --- | --- |
| 1st | 2.24 | $8 \pm 5$ |
| 1st | 2.09 | $24 \pm 6$ |
| 2nd | 2.85 | $15 \pm 4$ |
| 2nd | 1.22 | $13 \pm 7$ |
| 3rd | 0.23 | $<50$ |
| 3rd | 1.73 | $28 \pm 5$ |

The distillation was performed with a fully saturated liquid feed at port LF2 at a process speed of FIC01 = 8.5 slpm, with 1.1% off-gas flow-rate and a reflux ratio of $R = 180$. The samples were drawn after the system was stable for six hours. The off-line analysis showed a concentration in the in-gas line of

$$c_F = (136 \pm 22) \text{ ppt}, \quad (12)$$

which was calculated from the average over three measurements.[6] For the liquid-out sample, portions of each of the three storage volumes were measured two times each, for a total of six measurements. The results are shown in Table 1.

The concentration of krypton in the purified liquid-out line is thus

$$c_B < 26 \text{ ppq} \quad \text{at 90\% confidence level (CL)}. \quad (13)$$

This is the first demonstration that cryogenic distillation is capable of producing xenon with impurities at the ppq-level. The krypton concentration is more than a factor 30 lower than ever achieved with the XENON100 column. Furthermore, it fulfills the requirements for XENON1T ($^{nat}$Kr/Xe $<$ 200 ppq) and for future multi-ton experiments like XENONnT, LZ or DARWIN, requiring a krypton content below 100 ppq [4–6]. Although the RGMS system is sensitive down to 8 ppq [10], only a limit on the concentration of the purified gas is given due to the spread in the different measurements. The actual output of the distillation plant might be even purer, but uncertainties in sample retrieval, storage and transport might lead to an increase of the krypton content over time. The reduction factor is conservatively estimated to be

$$\frac{c_F}{c_B} = F_{\text{red}} > 5.6 \cdot 10^3 \quad (90\% \text{ CL}). \quad (14)$$

This reduction was achieved with a stripping section of 55 cm. Applying the McCabe–Thiele diagram, as shown in Fig. 2, to the obtained numbers above, leads to 3.6 stages for the stripping section. Taking into account, that operating with a fully saturated liquid feed the input condenser does not count as an additional distillation stage and thus can be neglected, a HETP value for a single distillation stage of 15 cm or less can be derived.

## 5 Performance of the final distillation column

In order to increase the separation efficiency, the final setup (Phase-2) was extended to 2.8 m of package material with additional 3 possible feed ports LF4 to LF6, reaching a total height of about 5.5 m as shown in Fig. 6. Two additional improvements were made: It turned out that the Phase-1 column was not able to purify xenon with high concentrations of $\mathcal{O}(10 \text{ ppm})$ of helium, argon and krypton, or electronegative impurities like oxygen and water, due to thermodynamic instabilities. Thus, both condensers (input and top) were modified to improve the gas collection for the pro-cessing of highly contaminated xenon. Furthermore, a third differential pressure gauge was implemented to measure the liquid level inside the input condenser, to allow for a more precise monitoring of the xenon feed. After the successful construction and commissioning of the Phase-2 column at WWU Münster, the system was dismounted, shipped to the Gran Sasso underground laboratory LNGS and re-assembled underground.

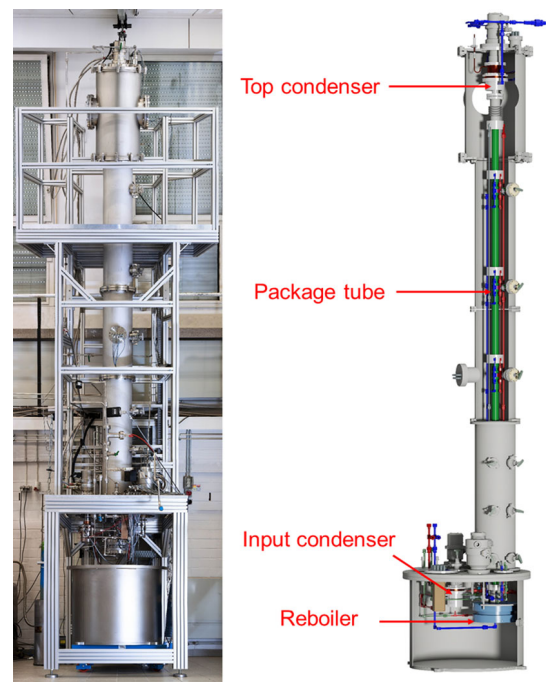

**Fig. 6** Phase-2 setup during commissioning at WWU Münster. The final height is about 5.5 m. The system was transferred and remounted at the underground laboratory LNGS in Italy

---

[6] This gas was already distilled before. Thus it exhibits a low krypton concentration.

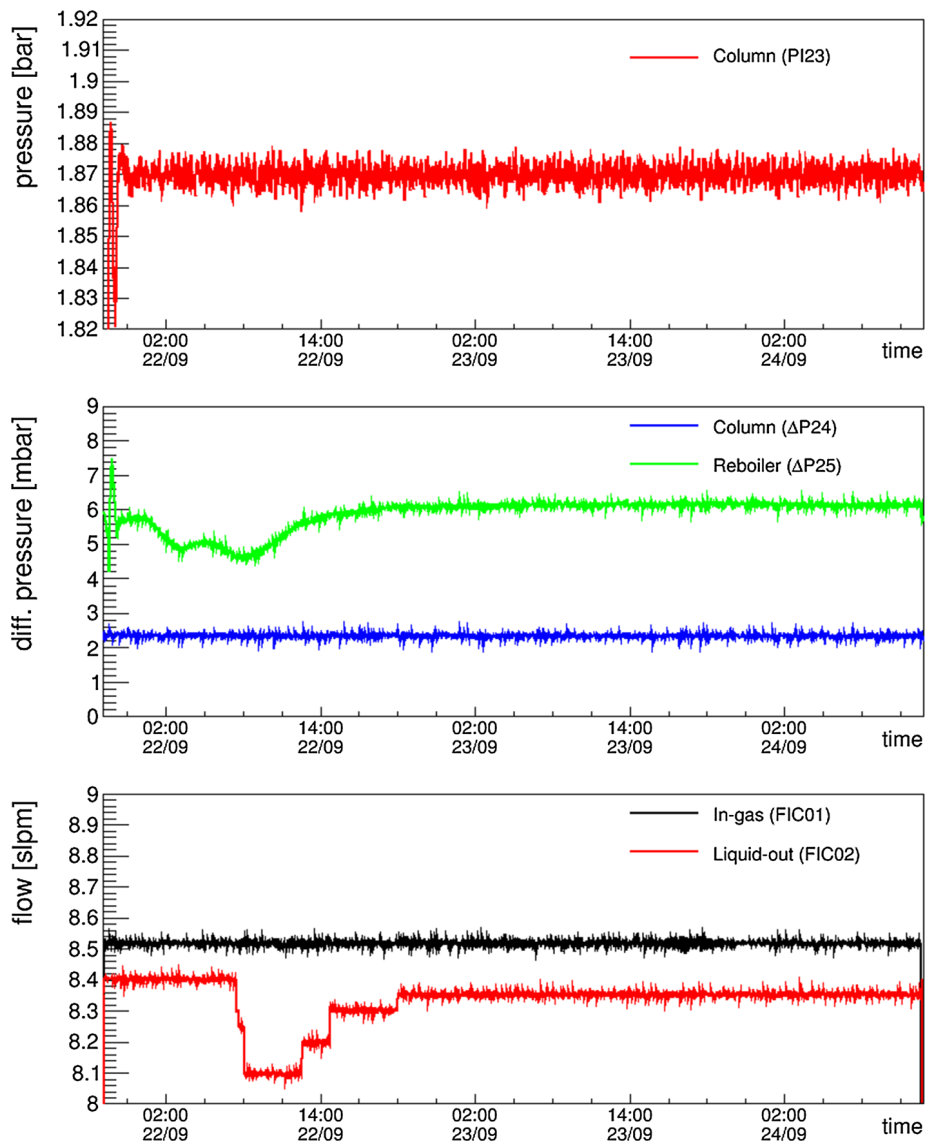

**Fig. 7** Thermodynamic stability of the final XENON1T distillation setup. The *plots* show the key parameters for a distillation test at LNGS over several days. The width and small spikes of the measurements are created by sensor inaccuracy and do not reflect the actual performance. *Top* Pressure PI23 inside the package material. *Middle* Pressure drop along the package tube ($\Delta$P24) and hydrostatic pressure of the liquid xenon reservoir in the reboiler ($\Delta$P25). *Bottom* Flow rates in the in-gas line (FIC01) and in the liquid-out line (FIC02). Thermodynamic stability is reached after about 24 h

### 5.1 Thermodynamic stability

Figure 7 shows the key parameters of the first long-term distillation run over several days at the Gran Sasso lab. About 212 kg of xenon were distilled with a fully saturated liquid feed at port LF5 at a flow rate of FIC01 = 8.5 slpm (Fig. 7, bottom) with 1% off-gas flow rate. The pressure stabilization works as efficiently as for the Phase-1 setup, stabilizing a few hours after starting the process, as shown in the top panel of Fig. 7. However, achieving a stable mass-balance is more complex since the Phase-2 setup has a longer package tube and the position of the feeding port, where the xenon is injected to the package tube, was also moved upwards. The overpressure inside the input condenser is therefore required to be much higher, to push the liquid upwards compared to the Phase-1 system. After about 24 h, the optimal settings were found and the liquid level inside the reboiler vessel ($\Delta$P25, Fig. 7, middle) remained constant at 6.2 mbar corresponding to a filling height of about 23 mm. The differential pressure along the package tube was measured to be about 2.4 mbar, corresponding to a pressure drop of about 0.86 mbar/m, similar to the Phase-1 system.

In further distillation studies at LNGS, higher flow rates of up to 16.5 slpm were operated stably while processing xenon with a krypton concentration of $\mathcal{O}(1\,\text{ppm})$. Furthermore, xenon with high concentrations of $O_2/Xe \approx 1300$ ppm and $N_2/Xe \approx 2600$ ppm was successfully processed. The studies indicate that the modifications of the gas collection in both input and top condenser were successful and allow processing xenon with a high content of impurities other than krypton.

### 5.2 Separation efficiency

During the commissioning run at WWU Münster, xenon artificially doped to a krypton concentration of $\mathcal{O}(10\,\text{ppm})$ was used to estimate the reduction factor of the Phase-2 column with the coldtrap-RGA method. Both the in-gas sample and the purified liquid-out gas were measured during the distillation process with 8.0 slpm. The concentration of natural krypton in the purified xenon was found to be below the sensitivity limit of the coldtrap-RGA system of $(206 \pm 36)$ ppt. The concentration in the supply bottle was determined to be $(20.2 \pm 5.4)$ ppm. This gives a reduction factor of

$$F_{\text{red}} > 9.4 \cdot 10^4 \quad (90\%\,\text{CL}), \tag{15}$$

proving the design value of $10^4$–$10^5$.

During the commissioning run at LNGS, 212 kg of xenon were processed (see Fig. 7), and several samples for measurements with the RGMS system were drawn. The in-gas was stored in four gas bottles. Three of them were measured with a commercial gas-chromatograph and have an average concentration of $(453 \pm 53)$ ppb. For the fourth bottle a certificate from the delivering company stated a concentration of less than 1000 ppb of krypton.

The sample container for the purified xenon was connected to the XENON1T purification system, which is designed as a gas manifold and allows extraction of samples. The purified liquid-out sample was measured to have a concentration of $(730 \pm 140)$ ppq. Assuming a uniform probability for the unknown concentration of the fourth bottle between 0 and 1000 ppb, the reduction factor is

$$F_{\text{red}} = (6.4^{+1.9}_{-1.4}) \cdot 10^5. \tag{16}$$

This reduction was achieved with a stripping section of 192 cm height. The resulting McCabe–Thiele diagram, adapted to Fig. 2, for the given numbers above, leads to 7 stages for the stripping section and thus to a HETP value for a single distillation stage of about 27 cm. This value is significantly higher than stated for the Phase-1 column. This may be evidence for reaching the limits of the description by the McCabe–Thiele method for ultra-low concentrations. Another possibility is, that systematic effects affecting sample collection, transport and preparation contaminated the liquid-out sample. The absolute concentration in the liquid-out line during the distillation process may be lower and thus the reduction factor might be even higher and the HETP value lower. In a forthcoming paper [23] we will present an independent determination of the HETP value using the $^{83m}$Kr tracer method [13,19].

**Table 2** Results on the krypton concentration measurements of different experiments

| Experiment | Technique | $^{\text{nat}}$Kr/Xe [ppq] |
|---|---|---|
| XENON100 | Distillation | 950 [10] |
| PandaX | Distillation | 1000 [11] |
| XMASS | Distillation | 2700 [24] |
| LUX | Chromatography | <200 [7] |
| XENON1T Phase-1 | Distillation | <26 |
| XENON1T final | Distillation | <48 |

Absolute concentrations <48 ppq (90% CL) were achieved with the final system when distilling xenon with input concentrations of about 50 ppb during further distillation runs. The achieved krypton concentrations are compared with other LXe-based dark matter experiments in Table 2.

## 6 Conclusion

A novel distillation column was designed and constructed for a process speed of 8.3 slpm (=3 kg/h) and a krypton reduction factor of $10^4$–$10^5$ to reach natural krypton in xenon concentrations <200 ppq. Although the design followed the McCabe–Thiele method, the development was done in two phases to ensure the validity of the construction for ultra-low concentrations. The first setup using 1.1 m of package material (Phase-1) was followed by the improved final setup (Phase-2), using 2.8 m of package material, currently installed in the underground laboratory LNGS as a subsystem of the XENON1T experiment. Both systems are thermodynamically stable at the design values. Higher process speeds of up to 18 slpm (=6.5 kg/h) were also shown to be stable.

The separation performance was investigated using different coldtrap RGA and RGMS techniques and the reduction factor of the final setup at LNGS has been measured to be $F_{\text{red}} = (6.4^{+1.9}_{-1.4}) \cdot 10^5$. The krypton concentrations delivered by the final system were measured to be <48 ppq, fulfilling the XENON1T requirement of <200 ppq. The lowest concentration achieved so far was <26 ppq, obtained with the Phase-1 column. This is almost an order of magnitude lower than needed for XENON1T.

The cryogenic distillation column is being operated at the XENON1T experiment and allows for processing 3.3 tons of xenon to produce an ultra-low krypton contamination at the ppq-level.

**Acknowledgements** We gratefully acknowledge support from the National Science Foundation, Swiss National Science Foundation, Deutsche Forschungsgemeinschaft, Max Planck Gesellschaft, German Ministry for Education and Research, Foundation for Fundamental Research on Matter, Weizmann Institute of Science, I-CORE, Initial Training Network Invisibles (Marie Curie Actions, PITNGA-2011-289442), Fundacao para a Ciencia e a Tecnologia, Region des Pays de la Loire, Knut and Alice Wallenberg Foundation, Kavli Foundation, and Istituto Nazionale di Fisica Nucleare. We are grateful to Laboratori Nazionali del Gran Sasso for hosting and supporting the XENON project. The Helmholtz Alliance for Astroparticle Physics supported S.R. and travel for I.C.